\documentclass[
    ,final            
  ]
  {aipproc}

\layoutstyle{8x11double}

\begin{document}

\title{Simulation studies of the high-energy component of a future
  imaging Cherenkov telescope array}


\author{S. Funk}{
  address={Kavli Institute for Particle Astrophysics and Cosmology,
    Stanford, CA-94025, USA}
}

\author{J.A. Hinton}{
  address={School of Physics \& Astronomy, University of Leeds, Leeds,
    LS2 9JT, UK}
}

\begin{abstract}
  The current generation of Imaging Atmospheric telescopes (IACTs) has
  demonstrated the power of the technique in an energy range between
  $\sim$ 100 GeV up to several tens of TeV. At the high-energy end,
  these instruments are limited by photon statistics. Future arrays
  of IACTs such as CTA or AGIS are planned to push into the energy
  range beyond 100 TeV. Scientifically, this region is very promising,
  providing a probe of particles up to the 'knee' in the cosmic ray
  spectrum and access to an unexplored region in the spectra of nearby
  extragalactic sources. We present first results from our simulation
  studies of the high-energy part of a future IACT array and discuss
  the design parameters of such an array.
\end{abstract}

\maketitle


\section{Introduction}

At the highest energies accessible to current Cherenkov telescope
arrays ($\sim$ 50 TeV) the sensitivity limit is imposed by the small
number of gamma-ray events which are collectible in reasonable ($\sim$
100 hours) exposures. The obvious method to increase sensitivity in
this important energy domain is therefore to build an instrument with
a collection area much larger than the $\sim10^{5}$ m$^{2}$ of IACT
arrays such as H.E.S.S.\ and VERITAS. High energy arrays have been
proposed in the past with the goal of achieving $10^{7}$ m$^{2}$
(i.e.\ 10 km$^2$) collection areas~\citep{Rowell:10TeV}. However, so
far it is far from clear what the most cost effective
telescope/camera/array layout for such an system would be.  We have
started investigations into the performance of a series of telescope
arrays with different telescope dish sizes and field of view / pixel
size with the aim of finding an optimal solution for achieving
reasonable angular and energy resolution over a huge area - at
manageable cost. This work is particularly relevant to the design of
the high energy part of future ground-based arrays such as AGIS and
CTA.

\begin{figure}
  \includegraphics[width=.48\textwidth]{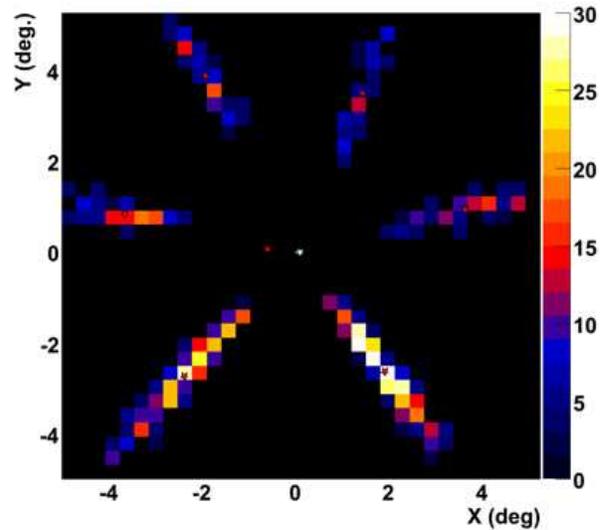}
  \caption{A simulated 14~TeV gamma-ray shower seen in the
    10$^{\circ}$ cameras of 6 out of the 7 telescopes of a sub-array
    of a high energy IACT system (the central telescope image is
    removed for clarity), as an illustration of shower imaging at high
    energy and large impact distance.  The centroid of each image is
    marked by a red star and the reconstructed (true) direction marked
    with a white star (green cross). The z-scale is in units of
    photoelectrons/pixel.}\label{fig::Fig1}
\end{figure}

\section{Design Considerations}

\subsection{Field of view}

\begin{figure}[h]
  \includegraphics[width=.48\textwidth]{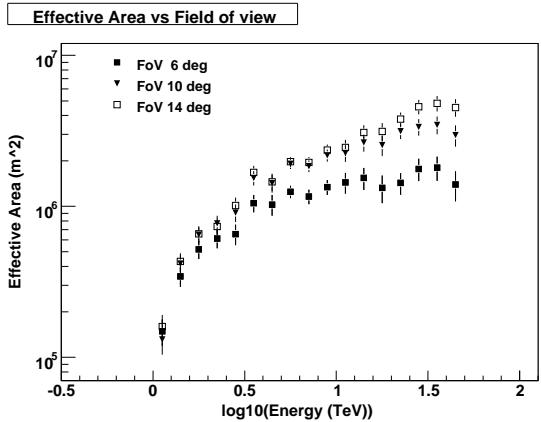}
  \caption{Effective area versus energy for three different field of
    views. For these studies, the pixel size has been fixed to 0.3°
    with seven 6~m-dishes on a triangular grid of 500~m spacing. 
    It can clearly be seen that at high energies a larger field of view 
    corresponds directly to a larger effective collection area.}\label{fig::Fig2}
\end{figure}

As a significant amount of Cherenkov light arrives on the ground at
very large distances from the shower axis (i.e. well beyond the $\sim$
100 m radius of the nominal ``Cherenkov light pool"), the collection
area of a single IACT (at the trigger level) is determined in the high
energy limit only by the camera field of view (FoV). Current systems
with $\sim 4^{\circ}$ FoV are sensitive only to showers with impact
distances of $\leq 200$m and hence reach asymptotic areas of $\sim
10^5 $ m$^2$. To achieve a 10 km$^2$ collection area such an array
would require at least 100 such telescopes. The alternative of fewer,
more widely spaced telescopes with wider FoV is clearly attractive,
but this approach is also challenging. Wide FoV systems may suffer
from problems with optical aberrations, camera weight and
obscuration. Detailed studies are needed, however, at this point it
seems likely that the cost of telescopes may rise rather quickly
beyond the 10$^{\circ}$ FoV suggested by~\citep{Biller:WideFoV} and offset
the benefits of wider telescope separation and/or increased telescope
multiplicity in individual events. As an example of how the effective
area changes with increasing FoV see Figure~\ref{fig::Fig2}.

\subsection{Pixel Size}
At large impact distance, air-shower images become rather long (see for
example Figure~\ref{fig::Fig1}), and the orientation of the image (and
hence the shower direction for multiple images) can be determined
reasonably well even with rather coarse (e.g. 0.3$^{\circ}$)
pixels~\citep{Biller:WideFoV}. However, the requirement to reject the
hadronic background implies that pixel sizes much larger than the rms
\emph{width} of gamma-ray showers will result in a rapid decrease in
sensitivity. Our studies have shown that there is a modest increase in
angular resolution with smaller pixels at TeV
energies~\citep{Funk:AngularResolution}. However, since for a fixed FoV
a decrease of a factor of a few in pixel size implies an order of
magnitude increase in cost, it seems advantageous to utilise as large
pixel sizes as possible (taking into account the rms width of
gamma-ray showers as mentioned above). We note that only vertical showers 
are simulated here: the reduction in angular width of showers at larger 
zenith angles may necessitate the use of somewhat smaller pixels for 
a realistic observing program.

\subsection{Telescope size}
Large telescopes are unavoidably required to reach low energies and
hence it has often been assumed that small telescopes are best for
high energy studies (for reasons of cost-effectiveness). There are, however, two
reasons for considering larger ($\sim 100$m$^2$) telescopes for
studies beyond 10 TeV. Firstly, at a fixed energy the impact distance
range of such telescopes is larger and at a fixed impact distance and
energy, higher amplitude signals are recorded - potentially resulting in improved
angular resolution and background rejection. Secondly, the cost of
a wide FoV camera is considerable (even with large pixels) and it is
unlikely that a design where the total cost of a telescope is
completely dominated by the camera cost is optimum, i.e. larger
telescopes could be used with little impact on the total cost of the
array. Therefore, whilst at first glance a 
widely spaced array of small telescopes ($<40$ m$^{2}$) with large
fields of view and large pixels seems the best approach 
for a cost-optimised high-energy array, detailed studies of 
performance versus cost are needed to find the optimum array 
parameters. In the following we present the first steps in such 
a study.

\section{Configurations}
For our initial work we considered sub-arrays containing 3 or 7
telescopes -- envisaged as components of a larger system. These arrays
are arranged on a triangular grid with separations of both 300~m and
500~m. The CORSIKA air-shower program~\citep{Corsika} is used together
with a simplified telescope simulation/ analysis program based on
idealised telescopes with square cameras and pixels. Vertical showers
are considered for simplicity. Telescope diameters of 6~m and 10~m are
considered. We estimate that the cost of a system of three 10~m
telescopes with 0.2$^{\circ}$ pixels is similar to that of a seven
telescope system with 0.3$^{\circ}$ pixels (both systems have a
10$^{\circ}$ FoV). The performance of these alternative systems are
compared in the following section, i.e.\ a system with a few large
telescopes and a system with more, but smaller, telescopes. An example
event as seen by the (500~m separation) 6~m system is shown in
Figure~\ref{fig::Fig1}, illustrating the necessity of a large FoV for
such impact distances, and that reasonable images can be obtained at
such energies and impact distances even with large (0.3$^{\circ}$)
pixels.

\section{First Results on the performance}

In a first step the effective areas for the two systems (3 large 10-m
telescopes with 0.2$^{\circ}$ pixels and 7 smaller 6-m telescopes with
0.3$^{\circ}$ pixels) both with 10$^{\circ}$ fields of view have been
compared. 

\begin{figure}[h]
  \includegraphics[width=.48\textwidth]{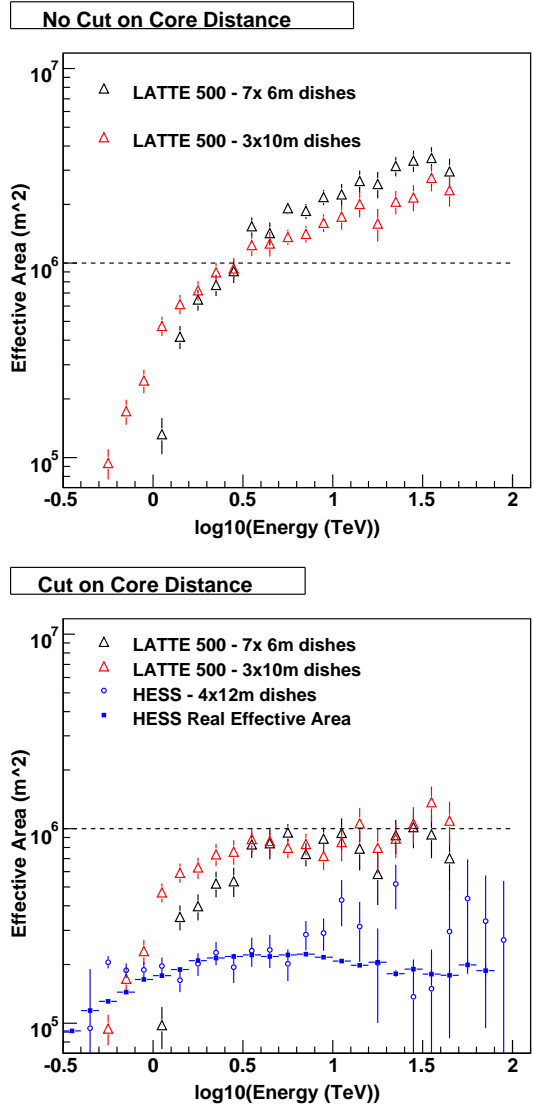}
  \caption{Effective area as a function of energy for the array with
    500~m telescope separation and a field of view of
    10$^{\circ}$. The red points show the effective area for a cell of
    3 of the larger telescopes (10~m dishes) with smaller pixels
    (0.2$^{\circ}$), the black points the effective area for a cell of
    7 telescopes with smaller (6~m) dishes and larger pixels
    (0.3$^{\circ}$). The top plot shows the effective area using all
    events, the bottom plot using only events within 500~m of the
    center of the array. Also shown in the bottom plot is the
    H.E.S.S.\ effective area at zenith.}\label{fig::Fig3}
\end{figure}

Figure~\ref{fig::Fig3} shows the resulting collection areas,
demonstrating that even with these rather inexpensive systems a
pre-cut effective area of well beyond $10^{6}$m is possible (see top
plot for the pre-cut effective areas). Cutting on the impact distance
of the showers, i.e.\ using only those showers with reconstructed core
position within 500~m from the center of the array, the effective area
still reaches $\sim 10^6$m (see Figure~\ref{fig::Fig3} bottom). As expected,
the system of large telescopes performs better at the low energies
(more light in the camera, therefore lower energy threshold), whereas
at the high-energy end, the system of more small telescopes gives a
larger effective area (due to the larger geometric area of the array). The lower
panel of Figure~\ref{fig::Fig3} also shows the H.E.S.S.\
effective area curve (in blue), open circles are derived using our simplified scheme, 
filled squares as determined using a detailed telescope simulation,
demonstrating the validity of our approach. It is evident from
these curves, that even with the modest $7 \times 6$m dish-array an
improvement in effective area of a factor of $\sim$5 over H.E.S.S.\ is
possible. This effect is due both to the larger area covered by
telescopes and to the increased field of view.

\section{Angular Resolution}
\begin{figure}[h]
  \includegraphics[width=.48\textwidth]{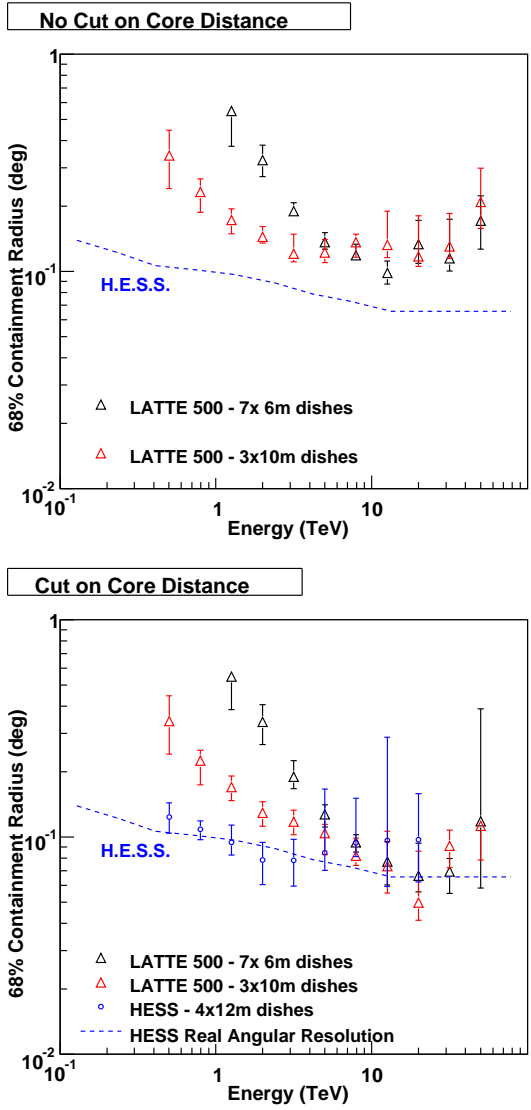}
  \caption{Angular resolution as a function of energy without cuts on
    impact distance (top) and with a cut of reconstructed core within
    500~m from the center of the array. For comparison the H.E.S.S.\ post-cut angular
    resolution is shown in blue - dashed for the H.E.S.S.\ array as
    determined from the full MC chain, blue circles from our simplified
    approach.}\label{fig::Fig4}
\end{figure}

The next step in characterising the high-energy array performance 
is to assess the angular resolution for the two configurations,
as shown in Figure~\ref{fig::Fig4}. The H.E.S.S.\
angular resolution is again shown for comparison and to demonstrate the
validity of our approach. As expected, the $7 \times 6$m array shows a
poorer angular resolution than the array with the larger dishes and
smaller pixels (basically due to the coarser pixelisation). In the
bottom plot a cut on the reconstructed core distance has again been
applied to select better reconstructed events. It should be noted,
that these results are preliminary and the angular resolution will likely
improve with a better adapted analysis method. Also evident from
Figures~\ref{fig::Fig3} and \ref{fig::Fig4} is that a real
optimisation of the array is needed since there are obvious trade-offs
between a larger effective area at the highest energies and the
angular resolution of the system.  It is evident from the comparison
of the bottom plots in Figures~\ref{fig::Fig3} and \ref{fig::Fig4}
that with a simple cut on the impact distance of the shower, the
angular resolution can be improved at the expense of a reduction in
effective area.

\section{Summary}
This study is a first attempt to study how to optimise array
parameters of the high-energy component of a future ground-based TeV
gamma-ray instrument such as AGIS or CTA. It is evident from the results presented
here that there are trade-offs between a large effective area and a good angular
resolution and a careful optimisation (with input on \emph{science} performance) 
has to be made to establish the optimum array design. In a next step, we will 
begin and investigation of background rejection power as a function of the primary
design parameters of telescope size, field of view and pixel size.

\begin{theacknowledgments}
  The authors would like to thank the members of the H.E.S.S.\, CTA
  and AGIS collaborations for help and interesting discussions. In
  particular we would like to acknowledge the help of K. Bernl{\"o}hr
  in all issues related to the MC production.
\end{theacknowledgments}



\bibliographystyle{aipproc}   

\bibliography{Gamma2008_LATTE}

\IfFileExists{\jobname.bbl}{}
 {\typeout{}
  \typeout{******************************************}
  \typeout{** Please run "bibtex \jobname" to optain}
  \typeout{** the bibliography and then re-run LaTeX}
  \typeout{** twice to fix the references!}
  \typeout{******************************************}
  \typeout{}
 }

\end{document}